\documentclass[12pt]{iopart}
\usepackage{amssymb}
\usepackage{graphicx}   
\usepackage{verbatim}   
\usepackage{color}      
\usepackage{graphics}
\usepackage{url}
\usepackage[numbers]{natbib}
\usepackage{units}
\usepackage{textcomp}
\usepackage[normalem]{ulem}
\usepackage[utf8]{inputenc}
\usepackage[T1]{fontenc}

\begin{document}

\title[Temp. charact. of HHG by THz Streaking]{Temporal characterization of individual harmonics of an attosecond pulse train by THz streaking}

\author{F. Ardana-Lamas$^{1,2}$, C. Erny$^{2}$, A. Stepanov$^{1}$, I. Gorgisyan$^{2}$, P. Juranic$^{2}$, R. Abela$^{2}$ and C.P. Hauri$^{1,2}$}
\ead{fernando.ardana@psi.ch;christoph.hauri@psi.ch}
\address{$^{1}$ Paul Scherrer Institute, CH-5232 Villigen PSI, Switzerland }
\address{$^{2}$ \'{E}cole Polytechnique F\'{e}d\'{e}rale de Lausanne, 1015 Lausanne, Switzerland}

\begin{abstract}

We report on the global temporal pulse characteristics of individual harmonics in an attosecond pulse train by means of photo-electron streaking in a strong low-frequency transient. The scheme allows direct retrieval of pulse durations and first order chirp of individual harmonics without the need of temporal scanning. The measurements were performed using an intense THz field generated by tilted phase front technique in LiNbO$_3$. Pulse properties for harmonics of order 23, 25 and 27 show that the individual pulse durations and linear chirp are decreasing by the harmonic order. 
 
\end{abstract}

\maketitle

\section{Introduction}

High order harmonic generation (HHG) \cite{lhuillier_high-order_1993} is an established technique for the formation of attosecond pulses \cite{paul_observation_2001,lopez-martens_amplitude_2005} which have been used for recording inner atomic electron dynamics on the sub-femtosecond time scale \cite{mauritsson_coherent_2008}. The HHG process ejects an attosecond burst every half cycle of the driving laser field which results in the formation of an attosecond pulse train (APT) as consequence of the coherent summation of individual bursts. In the frequency domain this translates to a regularly separated harmonic comb which eventually goes up to the keV photon energy range \cite{popmintchev_bright_2012}. 

Recently there is a raising number of frequency-selective applications which employ only one single individual harmonic of the comb. Examples include free electron laser seeding \cite{erny_metrology_2011,lambert_injection_2008,ardana-lamas_spectral_2013}, probing of ultrafast magnetic dynamics at the absorption edges \cite{lambert2015towards,heidler2015manipulating,kfir_generation_2014}, as well as time- and angle-resolved photoemission spectroscopy (tr-ARPES) \cite{eich_time-_2014}. Those dynamics have a corresponding frequency bandwidth similar to the bandwidth of a specific individual harmonic. Thus for resonant processes metrology on an individual harmonic in time turns therefore essential. 

Here we present a temporal characterization of individual harmonics using the streaking technique \cite{fruhling_single-shot_2009,schutte_electron_2011,grguras_ultrafast_2012,juranic_scheme_2014} with a strong Terahertz transient. This scheme comes along with the advantage that recording can be performed single-shot on a time window matching the entire temporal envelope of the entire APT. Presently different scanning techniques are regularly used for temporal characterization of APTs, among them RABBIT (reconstruction of attosecond beating by interference of two-photon transitions) \cite{paul_observation_2001}, FROG-CRAB (Frequency-Resolved Optical Gating for Complete Reconstruction of Attosecond Bursts) \cite{mairesse_frequency-resolved_2005} and PROOF (Phase Retrieval by Omega Oscillation Filtering) \cite{chini_characterizing_2010}. While those scanning techniques provide information about the relative delay between the harmonics (attochirp) and average pulse duration of two neighboring  harmonics, information about the chirp and pulse duration of a single harmonic is not provided. Cross correlation techniques with an $\unit[800]{nm}$ pulse \cite{norin_time-frequency_2002,mauritsson_measurement_2004}, as well as the THz streaking scheme used here overcomes those shortages. Compared to the cross correlation approach the THz streaking is not a scanning technique.
While streaking with a few-cycle 800 nm pulse has been applied on a sub-femtosecond time window for reconstruction of an isolated attosecond pulse \cite{itatani_attosecond_2002}, the Terahertz streaking scheme is suitable for measuring the temporal shape of the individual harmonics and in prinicple as as of the entire, tens of femtosecond long APT. Terahertz streaking provides a quick characterization of the time structure of an individual harmonic as there is no need for scans. The scheme suits in particular the demands of low-repetition rate soft x-ray sources including plasma-based x-ray lasers, high-power HHG, SASE free electron lasers where single-shot temporal characterization is essential.  

\section{THz streaking}

The measurement of the pulse duration is based on the formalism presented for the femto- and attosecond streak camera \cite{constant_methods_1997,itatani_attosecond_2002}. The general principle of the streak camera is based on mapping the temporal characteristic of an X-ray beam onto the velocity distribution of an electron beam. In the following calculation atomic units are used unless specified. The energy distribution of the streaked electrons is given by \cite{itatani_attosecond_2002}:  
\begin{equation}\label{eq:spec_pe}
 S(W)=\left|\int_{-\infty}^\infty dt\, e^{i\phi(t)} \mathbf{d}_{\mathbf{p}}  \mathbf{E_H}(t)\exp{[i(W+I_p)\,t]}\right|^2 
\end{equation}
Where $d_{p(t)}$ is the dipole transition element, $E_H(t)$ the X-ray pulse to be characterized, W the energy of the generated photo electrons, and $\phi(t)$ the phase accumulated between the generation under the presence of an external streaking field 
and the detection in the spectrometer. 
Following \cite{mairesse_frequency-resolved_2005} and using the assumptions $U_p \ll W_0$ and a significantly shorter x-ray pulse than the optical cycle of the streaking field $(\tau \ll \frac{1}{\omega_{THz}})$, the phase term can be approximated to:
\begin{eqnarray}
 \phi(t) & \approx \sqrt{\frac{8\,W\, U_p}{\omega_{THz}^2}} \cos \theta \left(1- \frac{\omega_{THz}^2 t^2}{2}\right)
  =\frac{s}{\omega_{THz}^2}\left(1-\frac{\omega_{THz}^2t^2}{2}\right).
 \label{phase_eq}
\end{eqnarray}
Where $U_p=E_{0}^2/4\omega_{THz}^2$ is the ponderomotive energy of the electron in the streaking THz field with $E_{THz}(t)=E_{THz}(t)\cos(\omega_{THz}t)$ and the slope of the energy shift, the so called streaking speed $s=\frac{\partial \Delta W}{\partial t}=\sqrt{2W}E_{0}$ \cite{kienberger_atomic_2004}. The $\pm$ accounts the direction of the streaking field towards ($+$) or away ($-$) from the electron spectrometer.  

In the presented experiment the detection is in the plain of the streaking field with a narrow acceptance angle, therefore we assume $\theta = 0$. 
If we assume a Gaussian X-ray pulse
\begin{equation}
 E_H(t)=\exp\left\{\frac{-t^2}{4\tau^2}\right\}\exp\left\{-i(\omega_H t+ \xi t^2)\right\},
\end{equation}
with $\tau$ the rms pulse duration and $\xi$ the first order chirp. The
the length of the transform limited pulse with the same rms spectral width $\sigma_0$ is then given by
\begin{equation}
  \widetilde{\tau}^2=\frac{\tau^2}{1+16 \tau^4 \xi^2}=\frac{1}{4\sigma_0^2}.
\end{equation}
Under these conditions the integral of \eref{eq:spec_pe} can now be solved analytically and we obtain
\begin{eqnarray}
 S(W)\propto exp\left\lbrace \frac{-(W-W_0)^2}{2(\sigma_{0}^2 +\tau^2\,[s^2\pm 4\,|s| \,\xi])} \right\rbrace ,
 \label{eq:spectro}
 \end{eqnarray}
with $W_0=\omega_H-I_p$ the kinetic energy of the unstreaked photoelectrons.

As shown in \eref{eq:spectro} the broadening produced under the effect of the streaking field depends on the pulse duration as well as on the chirp. By combining the measurements with opposite streaking speed, e.g. opposite streaking direction, we can calculate directly both the pulse duration and the chirp using the following expressions: 
\begin{eqnarray}
\tau=\sqrt{\frac{\tilde{\sigma}_{+}^2+\tilde{\sigma}_{-}^2}{2 s^2}} \hspace{2cm}
\xi=\frac{\tilde{\sigma}_{+}^2-\tilde{\sigma}_{-}^2}{8s\tau^2}, 
\label{eq:final}
\end{eqnarray}
where $\tilde{\sigma}_{+/-}=\sqrt{\sigma_{+/-}^2-\sigma_0^2}$ is the de-convoluted rms spectral bandwidth under a positive/negative streaking.

To be able to retrieve the pulse duration a broadening at least equal to the x-ray bandwidth ($\frac{s \tau}{\sigma_{0}}\geq 1 \label{Eq:res}$, for unchirped pulses) needs to be observed \cite{itatani_attosecond_2002}. This condition settles the minimum pulse duration that can be retrieved. To be able to resolve shorter pulses higher streaking speed are necessary, however the approximations made in \eref{phase_eq} limits the maximum ponderomotive potential for which an analytical solution exists ($U_p \ll W_0$). In our experiment we estimate a ponderomotive energy of the THz field of $U_p \approx \unit[1.2]{eV}$, where as the measured kinetic energies of the photoelectrons are between 8 and \unit[20] {eV}, thus fulfilling the previous condition.

\section{Experimental setup}

\begin{figure}[floatfix]
	\centering
	\includegraphics[width=0.5\columnwidth]{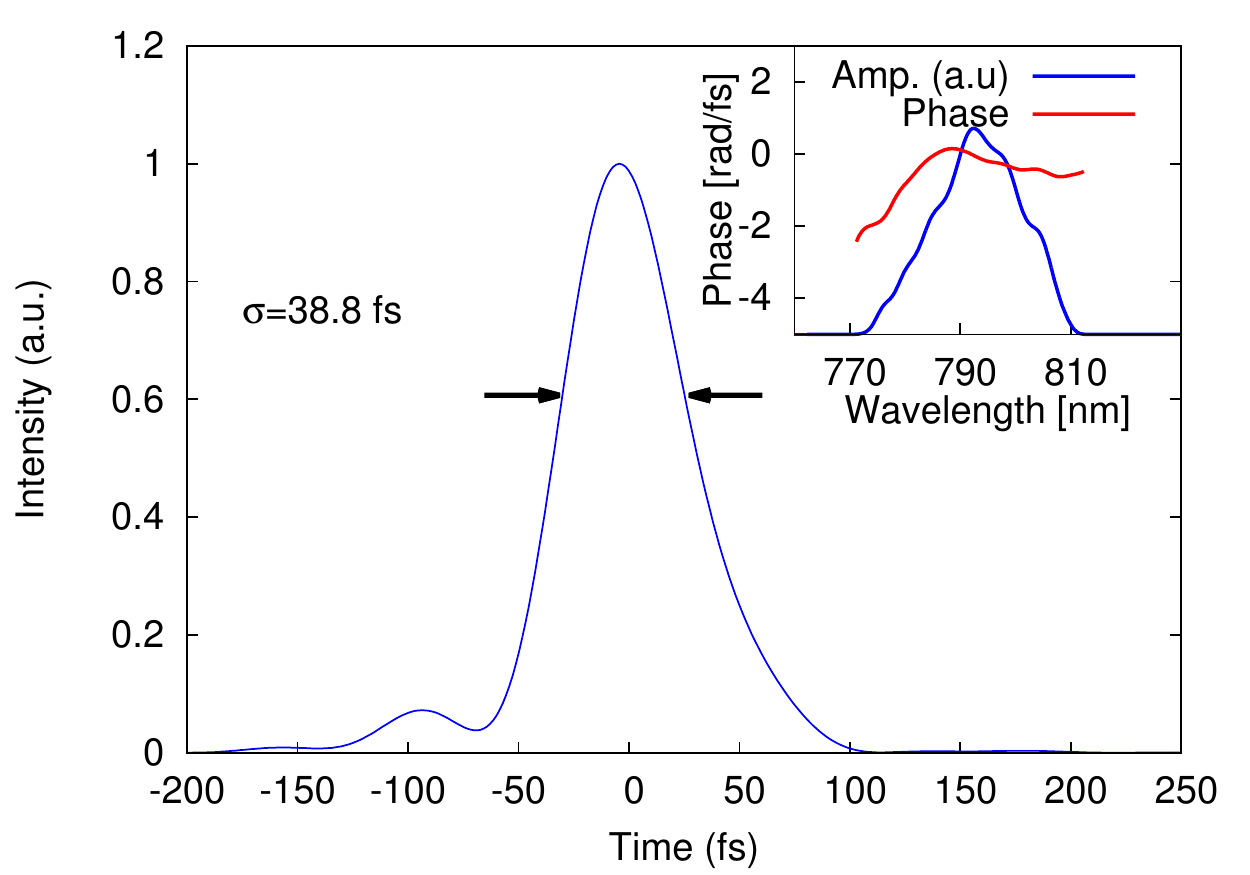}
	\caption{\label{fig:DrivingPulse}Temporal profile of the driving laser. The pulse length is \unit[38.8]{fs} (rms). Inset, spectral amplitude and phase of the driving pulse, with the latter optimized for most efficient high-order harmonics generation.}
\end{figure}

The experiment was performed using a Ti:Sapphire laser system (20 mJ, 100 Hz) with a pulse
duration of \unit[39]{fs} (rms) (\fref{fig:DrivingPulse}). The laser pulse energy was split to operate synchronously the THz source (60\%) as well as the high order harmonic generation (40\%).
 A loose focusing geometry (f=3 m) is employed for HHG at high flux \cite{ardana-lamas_spectral_2013}. The laser waist (85 $\mu$m, $\frac{1}{e^2}$ beam radius) was placed at the face of a 3 cm long unmodulated glass cell with 800 $\mu$m inner diameter. A solenoid-based valve situated at the center of the cell released bursts of argon with a backing pressure of up to 3 bar at the full repetition rate of the laser. Downstream the HHG beam is focused by a toroidal mirror (focal length of 4 m) into the detection zone of the streaking chamber installed at the end of the 6 meter long HHG beam line.  Two \unit[200]{nm} thick aluminum filters and a fused silica plate were used to separate the HHG beam from the infrared laser.
 
\begin{figure}[floatfix]
	\centering
	\includegraphics[width=0.5\columnwidth]{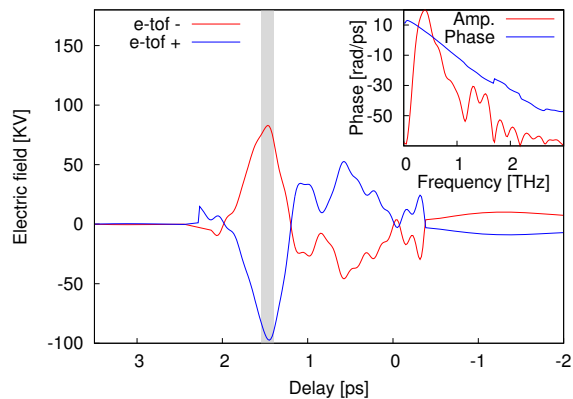}
 	\caption{\label{fig:THz_Desc_I}THz electric field obtained by recording photoelectrons produced by the short HHG pulse acting as probe. The field reconstructed by means of the e-tof left (right) is shown in red (blue).}
\end{figure}

THz radiation is produced directly next to the streaking chamber by optical rectification in an uncooled LiNbO$_3$ crystal. A tilted pulse front configuration was used to optimize for high power THz output \cite{stepanov_generation_2008}. This results in a ca. 10 $\mu$J pulse energy and a THz spectrum centered at \unit[0.3]{THz} (\fref{fig:THz_Desc_I}). After in-coupling into the streak camera the Terahertz radiation is overlapped collinearly with the HHG beam using a  parabolic mirror (f=\unit[170]{mm}) with a central hole. The latter was equipped with a \unit[5]{mm} central hole in order to let the HHG beam pass towards the interaction zone of the THz streak camera where both beams where focused and overlapped in time and space. Streaking field strength up to 100 kV/cm (\fref{fig:THz_Desc_I}) could be achieved at the gas jet position. 

\begin{figure}[floatfix]
	\centering 
	\includegraphics[width=0.95\columnwidth]{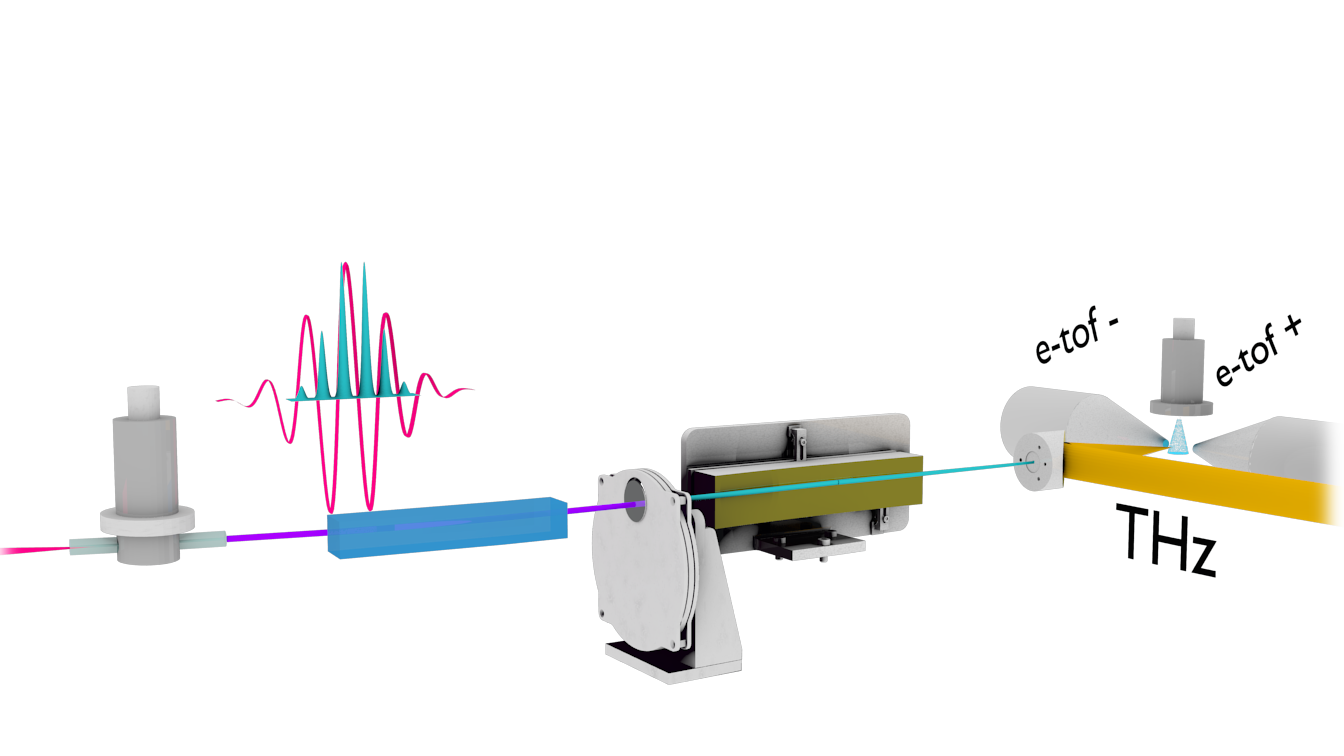}
	\caption{\label{fig:ExpLayout}Experimental layout with high-order harmonics generated in a capillary followed by harmonic separation (silicon plate and Al filters) and a 1:1 toroidal mirror (17 cm parent focal length) focusing the APT into the gas jet located in the detection zone of the two opposing time of flight electron spectrometers.}
\end{figure}

The streak camera (\fref{fig:ExpLayout}) consists of two electron time-of-flight (e-tof) spectrometers oriented parallel to the electric field and placed at opposing sides of the gas jet. For this configuration the recorded signals of the two e-tofs are complementary as it allows the simultaneous retrieval of positively and negatively streaked electrons (\fref{fig:THz_Desc_I}). The typical energy resolution of the e-tof was around 1\%. The photoelectrons were produced by single-photon ionization of atomic helium by the HHG beam. The gas was injected using a piezo controlled pulse valve with opening times typically around \unit[20]{$\mu$s} with a backing pressure between 1 and 4 bars.

\section{Experimental results}

\begin{figure}[floatfix]
	\centering
	\includegraphics[width=0.7\columnwidth]{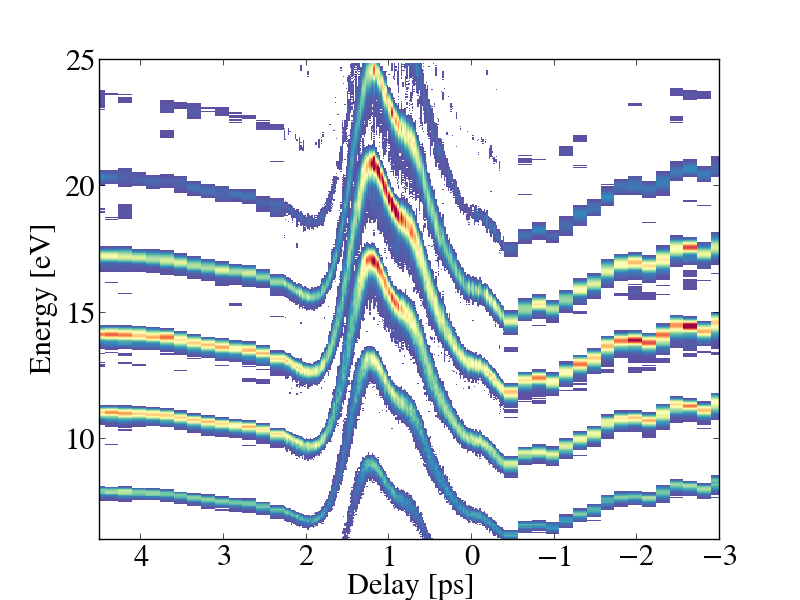}   
	\caption{\label{fig:Scan_1}Photoelectron spectra of individual harmonics generated in helium as function of the time delay between ATP and THz field. Higher order harmonics give rise to higher kinetic electron energies which leads to the observed energy separation of individual harmonics. In order to obtain a better resolution a finer time step was used around the peak of the THz pulse.}
\end{figure}

APT reconstruction requires knowledge about the THz streaking field. This characterization can be done either by conventional electro-optical sampling or, as we did it here, directly by recording the signal of the photoelectrons produced at different delays in the streak field (\fref{fig:Scan_1}). Our measurement unveiled a THz field asymmetry between the leading and the trailing edge. This indicates a small chirp of the streaking pulse which could originate from pulse distortion due to absorption in air or from the generation process itself. To avoid inconsistencies we evaluated only the data from the leading edge where the highest streaking speed and highest temporal resolution is achieved.

\begin{figure}[floatfix]
   \includegraphics[width=0.5\columnwidth]{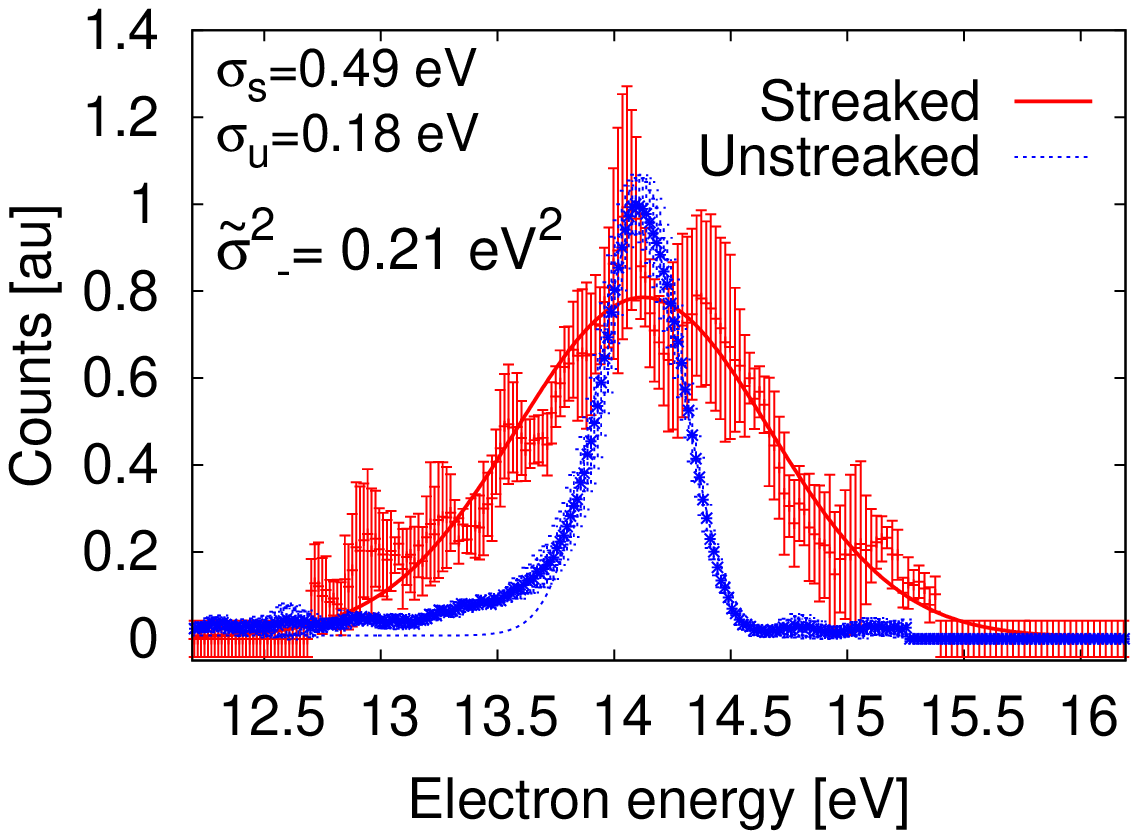}\hspace{-0.4cm}
   \includegraphics[width=0.5\columnwidth]{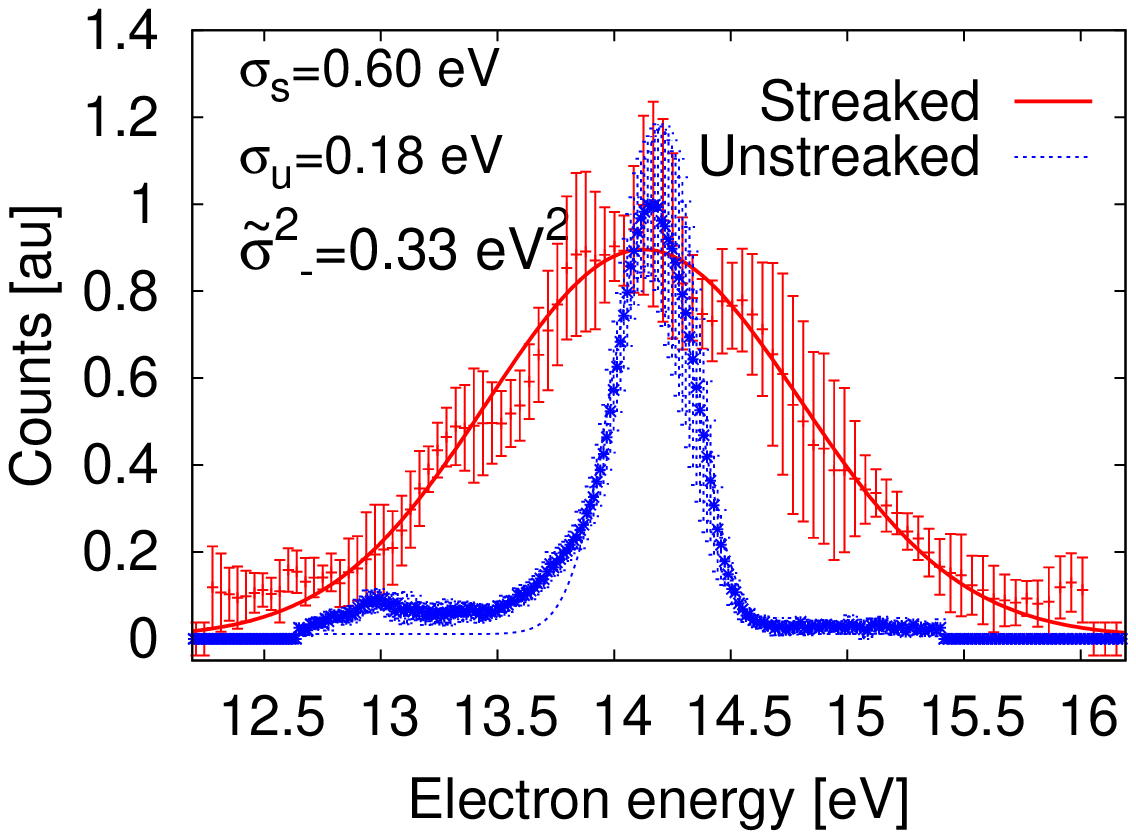}
  \caption{\label{fig:signals} Left (Right): 
  Comparison of the streaked and unstreaked photoelectron spectrum of harmonic 25 for the left  and right e-tof. The streaked signals have been shifted in offset in order to overlap with the unstreaked signal.}
\end{figure}

Photoelectron spectra for harmonic 25 are shown in (\fref{fig:signals}) for the streaked and unstreaked configuration.  A comparison of spectra recorded by the left and right e-tof, respectively, shows that the corresponding linewidths are differing slightly.  According to \eref{eq:final} this entails a linear chirp of about \unit[-0.9]{$\frac{meV}{fs}$}. 

\begin{figure}[floatfix]
	\includegraphics[width=0.5\columnwidth]{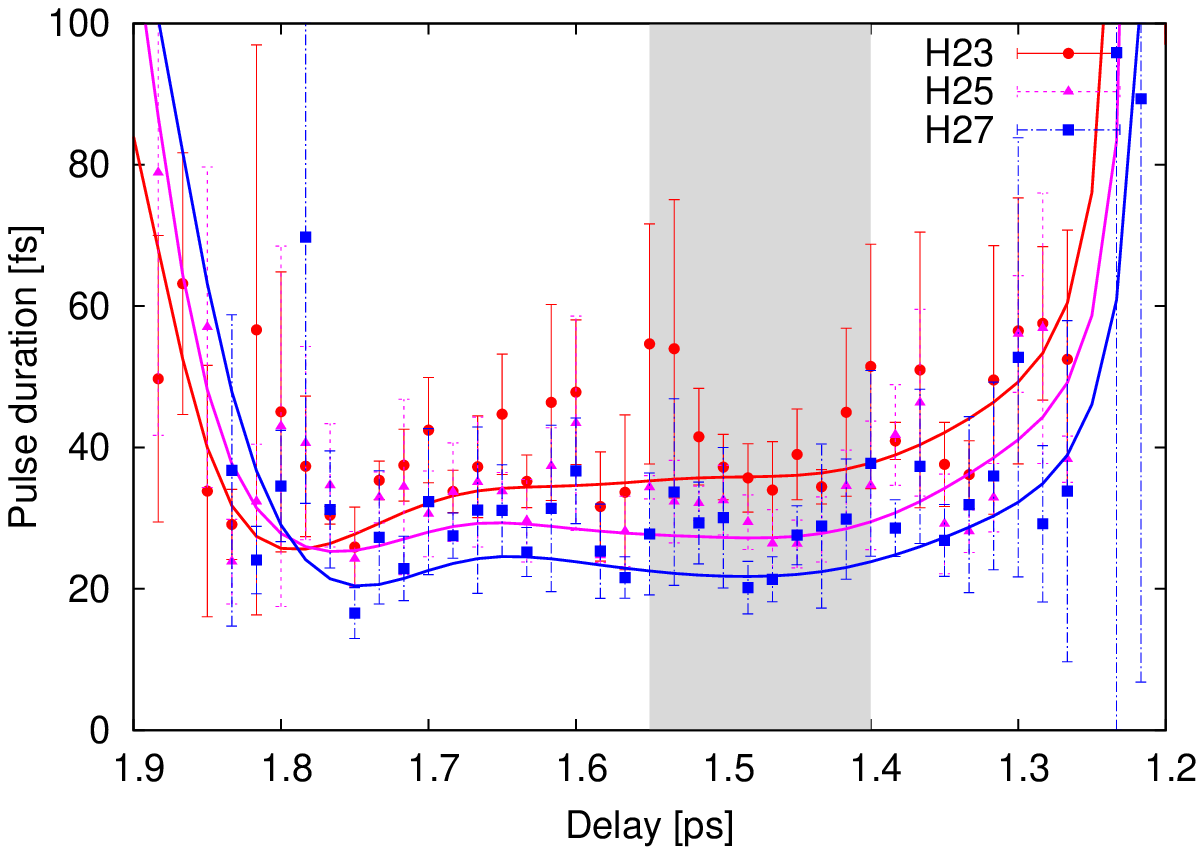}
	\includegraphics[width=0.5\columnwidth]{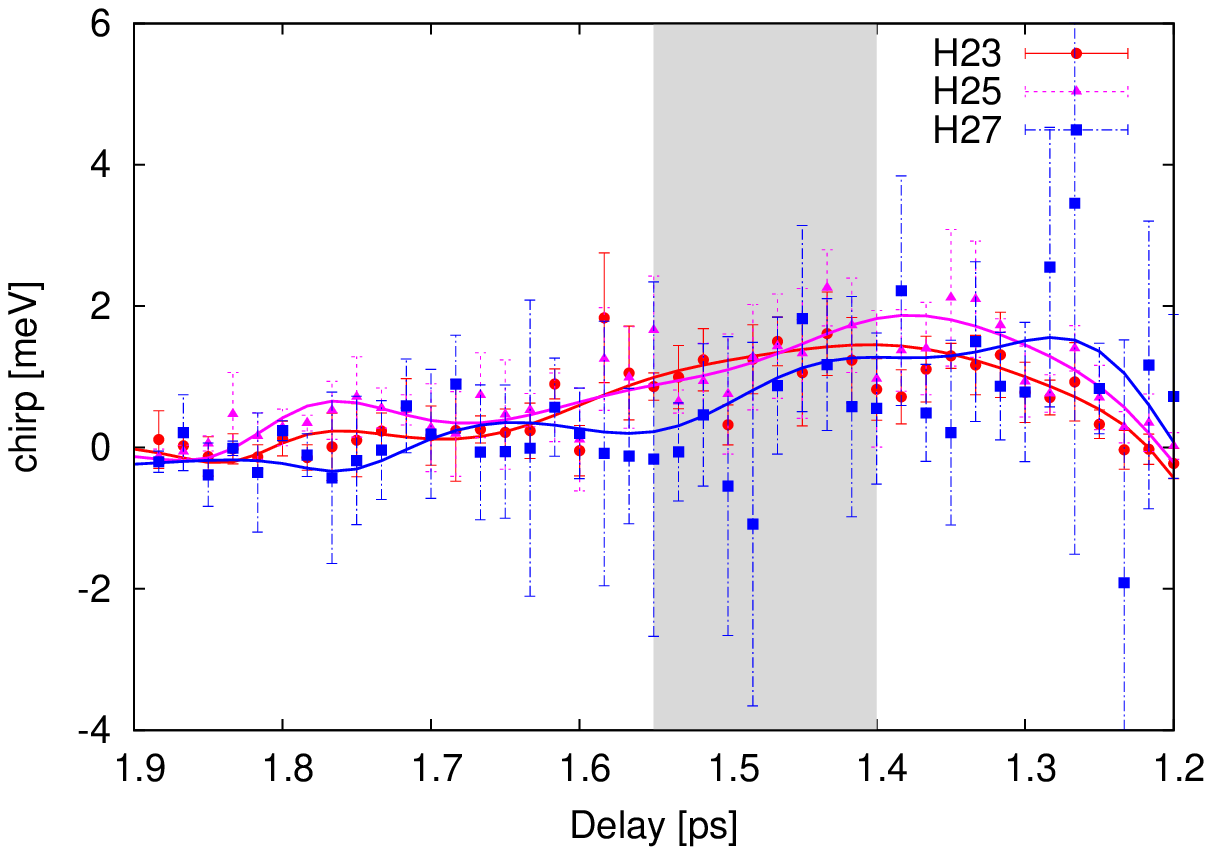}
	\caption{\label{fig:Scan_retrieved}Retrieved pulse duration (left) and chirp (right) as function of the time delay. Discrete points, averaged data over a set of measurements. Solid line represents the smoothed signal using a low-pass frequency filter. The grey area indicates the window where pulse reconstruction according to \eref{phase_eq} is valid. Pulse durations and chirp for individual harmonics are summarized in \tref{tab:T_Scan_1}. }
\end{figure}

For a more accurate retrieval of the pulse duration we have additionally performed a temporal scan around the first maximum of the electric field (corresponding to the linear part of the vector potential) with subsequent Fourier filtering to eliminate discontinuities between subsequent data points and to reduce the noise on the retreival of $W$ and $\sigma_{+/-}$. For each delay the pulse duration and chirp has been calculated. It can be observed (see \fref{fig:Scan_retrieved}) how the retrieved pulse duration is only valid in the region where the approximation made in \eref{phase_eq} holds (\fref{fig:Scan_retrieved}, grey area), e.g. the streaking speed deviates by less then 20 \% from the maximum. Outside this window the pulse characterization fails and the calculated pulse duration is clearly overestimated.  The pulse duration and chirp has been calculated as weighted average of the data inside of the evaluation window. The minimum pulse duration we are able to retrieve with the current configuration can be calculated by taking into account the linewidth of the unstreaked spectrum and the streaking speed.

In our case $\sigma_0= 0.18 \,eV$ and $s=0.02\,\frac{eV}{fs}$ support retrieval of a minimum pulse duration of about \unit[10]{fs} (rms). In these measurements, as expected from the highly nonlinear intensity-dependent HHG process, the pulse length for harmonics of increasing order (23, 25 and 27) is decreasing from 36 to 32 and 24 fs (rms), respectively (see \tref{tab:T_Scan_1}). This is in the agreement of the cut-off law of the HHG process, where the lower order harmonics can be generated at lower intensities, thus earlier in the leading edge and later in the trailing edge of the pulse, than the higher orders. The measured pulses show a clear negative chirp, which is in agreement with earlier measurements \cite{mauritsson_measurement_2004,varju_frequency_2005}.

Nevertheless, the full reconstruction of the APT requires the information on the delay between different harmonics, as e.g. measured by RABBIT. In the current configuration this requires either a CEP stable driver laser, or single shot acquisition with subsequent CEP binning. The CEP drifts lead to timing jitter in the order of the laser period, compared to the THz field and is therefore averaged out by this measurement technique. 
We finally mention  that the streaked and unstreaked photoelecton spectra are affected by the space charge forces producing an artificial broadening. Nevertheless, the streaking technique is not affected by space charge effects because the pulse reconstruction relies on the difference signal between the streaked and unstreaked spectrum, canceling out the effect of space charge effects which are present in both cases.

\begin{table}[floatfix]
\centering
  \begin{tabular}{c|c|c|c}
   \bfseries Harmonic & \bfseries Pulse length & \bfseries Chirp & \bfseries Transform limit \\
   \bfseries [-] & \bfseries  (rms) [fs] & \bfseries [meV/fs] & \bfseries (rms) [fs]  \\
   \hline
   23 & $36 \pm 5$ & $ -0.96 \pm 0.08$ & $4.7 \pm 0.7$  \\
   \hline
   25 & $32 \pm 4$ & $ -0.91 \pm 0.1 $ & $6 \pm 1$ \\
   \hline
   27 & $24 \pm 2$ & $ -0.82 \pm 0.1 $ & $8 \pm 1$
  \end{tabular}
  
  \caption{\label{tab:T_Scan_1}Summarized results from \fref{fig:Scan_retrieved}. The pulse duration and chirp is calculated as the weighted average over the region where the approximation $\cos \omega_{THz} t \approx 1- \frac{\omega_{THz}^2 t^2}{2}$ is valid (grey area).}
\end{table}

\section{Conclusion}

In summary we have shown that streaking of photoelectrons by a low-frequency Terahertz transient is suitable to measure the pulse duration of the individual harmonics of an attosecond pulse train. We have characterized the individual pulse length and chirp associated to the harmonics of order 23 to 27 generated in a loose-focusing configuration. The individual harmonic pulses are shown to present a linear chirp and pulse durations that decrease with the harmonic order. The transient streaking technique explored here is useful for experiments where the temporal profile of an individual harmonic is required, and for single-shot pulse shape characterization of sources where conventional pulse reconstruction schemes based on scanning are not applicable. 

\section{Acknowledgments}
\textit{CPH and FA acknowledge financial support
from the Swiss National Science Foundation under grant PP00P2\_128493, SwissFEL and COST-MP1203 under contract C13.0116. CPH acknowledges also association to NCCR-MUST. }

\bibliography{References_ce}
\bibliographystyle{unsrt}

\end{document}